\documentclass[twocolumn]{aastex631}
\newcommand{\hi}{H\textsc{i}}
\newcommand{\hii}{H\textsc{ii}}
\newcommand{\fcnm}{$f_{\rm CNM}$}
\newcommand{\fpah}{$f_{\rm PAH}$}
\newcommand{\qpah}{$q_{\rm PAH}$}

\newcommand{\rhi}{$\mathcal{R}_{\rm \hi}$}
\providecommand{\sorthelp}[1]{}

\begin{document}

\title{Polycyclic Aromatic Hydrocarbons, the Anomalous Microwave Emission, \\ and Their Connection to the Cold Neutral Medium}

\author[0000-0001-7449-4638]{Brandon S. Hensley}
\email{bhensley@astro.princeton.edu}
\affiliation{Department of Astrophysical Sciences,  Princeton University, Princeton, NJ 08544, USA}
\affiliation{Spitzer Fellow}

\author[0000-0002-7743-8129]{Claire E. Murray}
\affiliation{Space Telescope Science Institute, 
3700 San Martin Drive, 
Baltimore, MD, 21218}

\author[0000-0002-3352-9272]{Mark Dodici}
\affiliation{Department of Astrophysical Sciences,  Princeton University, Princeton, NJ 08544, USA}

\date{\today}

\begin{abstract}
Using new large area maps of the cold neutral medium (CNM) fraction, $f_{\rm CNM}$, we investigate the relationship between the CNM, the abundance of polycyclic aromatic hydrocarbons (PAHs), and the anomalous microwave emission (AME). We first present our $f_{\rm CNM}$ map based on full-sky HI4PI data, using a convolutional neural network to covert the spectroscopic \hi\ data to $f_{\rm CNM}$. We demonstrate that $f_{\rm CNM}$ is strongly correlated with the fraction of dust in PAHs as estimated from mid- and far-infrared dust emission. In contrast, we find no correlation between $f_{\rm CNM}$ and the amount of AME per dust emission, nor between $f_{\rm CNM}$ and the AME peak frequency. These results suggest PAHs preferentially reside in cold, relatively dense gas, perhaps owing to enhanced destruction in more diffuse media. The lack of correlation between $f_{\rm CNM}$ and AME peak frequency is in tension with expectations from theoretical models positing different spectral energy distributions of AME in the cold versus warm neutral medium. We suggest that different PAH abundances and emission physics in different interstellar environments may explain the weaker than expected correlation between 12\,\micron\ PAH emission and AME even if PAHs are the AME carriers.
\end{abstract}


\section{Introduction}
Polycyclic aromatic hydrocarbons (PAHs) are among the smallest interstellar grains and account for only $\sim5\%$ of the total dust mass \citep{Draine+Li_2007}. Nevertheless, PAHs dominate the mid-infrared (MIR) spectrum of galaxies via emission in discrete vibrational bands between 3 and 20\,\micron\ \citep{Leger_1984,Allamandola_1989}. Indeed, it is typical for 10--20\% of the bolometric infrared emission of star-forming galaxies to be emitted by PAHs \citep{Smith+etal_2007}. In part due to this brightness, PAH emission has emerged as a powerful tracer of star formation \citep[e.g.,][]{Shipley+etal_2016,Xie+Ho_2019,Lai+etal_2020}.

Resolved photometry of nearby galaxies shows enhancement of the fraction of dust mass in PAHs in diffuse and molecular gas relative to \hii\ regions \citep{Chastenet+etal_2019}. In dense clouds, Galactic observations suggest that shielding from UV radiation allows PAHs to grow to larger sizes and even convert to non-PAH species, such as fullerenes, via coagulation and chemical processing \citep[e.g.,][]{Berne_2012,Croiset_2016}. The shift in PAH size to larger molecules coupled with the lack of exciting photons yields a reduction in PAH emission \citep[e.g.,][]{Draine_2021}. In regions of such high volume density, PAHs can stick to the surfaces of larger grains and thus no longer attain the temperatures required to emit in the MIR. In more diffuse gas, the smallest grains are rapidly sublimated in the interstellar radiation field \citep{Guhathakurta_1989}. PAHs exposed to harsh UV radiation in photoionized gas may be destroyed \citep{Dong_2011}, and in hot gas grains can be eroded by sputtering \citep{Draine1979}. This supports a picture in which PAH emission arises preferentially in diffuse atomic to moderately dense molecular gas, but is suppressed both via grain destruction in the most diffuse gas and via coagulation and lack of exciting photons in dense molecular environments.

The anomalous microwave emission (AME) is dust-correlated emission peaking near 30\,GHz that greatly exceeds extrapolation of the far-infrared (FIR) dust emission to these frequencies \citep[see][for a review]{Dickinson_2018}. The leading hypothesis for the origin of this emission is rotational electric dipole emission from sub-nanometer interstellar grains \citep{Draine_1998}, i.e., ``spinning dust emission.'' As PAHs are ubiquitous in the Galactic interstellar medium (ISM) and are generally expected to have non-zero electric dipole moments, they have provided a natural hypothesis for the identity of the AME carriers \citep{Draine_1998}.

Observational evidence for a link between PAHs and AME is mixed. On large angular scales, it is well-established that PAH emission and AME are strongly correlated \citep[e.g.,][]{Planck_Int_XLVIII,Hensley_2016,Dickinson_2018}. The question, however, is whether this correlation is any stronger than that between FIR dust emission and the AME, as might be expected if PAHs are the AME carriers. In the Perseus molecular cloud and the \hii\ region RCW175, the PAH abundance is found to be poorly correlated with AME on arcminute scales \citep{Tibbs_2011, Tibbs_2012}. On the other hand, a recent study of the $\rho$ Oph molecular cloud on arcminute scales found that PAH emission is much better correlated with AME than is the FIR dust emission \citep{ArceTord_2020}. Degree scale studies of AME in $\lambda$ Orionis have found comparable levels of correlation between AME and either FIR emission or PAH emission \citep{Bell_2019,CepedaArroita_2021}. However, \citet{Bell_2019} found a significantly stronger correlation between PAH mass and AME than between total dust mass and AME in $\lambda$ Orionis. In the external galaxy NGC\,6946, the PAH mass surface density is no better correlated with AME than total dust surface density \citep{Hensley_2015}. In a sample of 14 protoplanetary disks, the three with significant AME had unremarkable PAH features \citep{Greaves_2018}.

The lack of strong correlation between PAH emission and AME has prompted consideration of other AME carriers besides PAHs, including nanosilicates \citep{Hoang_2016b,Hensley_2017,MaciaEscatllar_2020}, nanodiamonds \citep{Greaves_2018}, and Fe nanoparticles \citep{Hoang_2016a,Hensley_2017}. Any sufficiently abundant sub-nanometer species can in principle contribute to the AME as long as grains have a non-negligible electric or magnetic dipole moment.

Regardless of whether spinning dust emission arises from PAHs, its spectral energy distribution (SED) depends on the rotational excitation of grains and thus the local ISM conditions. In particular, theoretical calculations predict that the spinning dust emission arising from the cold neutral medium (CNM) peaks at higher frequencies than that from the warm neutral medium \citep[WNM;][]{Draine_1998,AliHaimoud_2009,Ysard_2010a,Hensley_2017}. When fitting multi-frequency AME SEDs in the diffuse ISM, two components are frequently needed to produce a sufficiently broad spectrum \citep{Ysard_2010b,Hoang_2011,Planck_2015_X}. These components have been associated with AME from the CNM and WNM \citep{Ysard_2010b,Hoang_2011}. This picture predicts that the AME spectrum should systematically shift to higher frequencies as $f_{\rm CNM}$ increases.

In this work, we undertake an empirical investigation of the interrelationships between PAHs, the AME, and the CNM. Our study is enabled by the availability of full-sky spectroscopic \hi\ data from HI4PI \citep{hi4pi2016} and new techniques for \hi\ phase decomposition. In particular, \citet{murray2020} trained a convolutional neural network (CNN) using synthetic observations of 3D hydrodynamic Galactic ISM simulations \citep{kim2013, kim2014} to predict the CNM fraction (\fcnm), a quantity which formally requires knowledge of \hi\ emission and absorption, from \hi\ emission at $21\rm\,cm$ alone. The model is applicable only at high Galactic latitudes ($|b|>30^{\circ}$), where the $21\rm\,cm$ synthetic training observations successfully reproduce characteristics of observed $21\rm\,cm$ emission and absorption spectra, which are relatively simple. \citet{murray2020} applied this model to the GALFA-\hi\ survey of $21\rm\,cm$ emission from the Arecibo Observatory \citep{peek2011, peek2018}, finding excellent agreement with constraints from available $21\rm\,cm$ absorption measurements. By extending the model from \citet{murray2020} to HI4PI, we probe the CNM across the full high-latitude sky ($|b|>30^{\circ}$). 

This paper is organized as follows: in Section~\ref{sec:data} we describe the data products used in this work; we present our new, full-sky map of $f_{\rm CNM}$ in Section~\ref{sec:fcnm}; we use this map to investigate the relationship between the CNM and PAH abundance in Section~\ref{sec:pahs}, and between the CNM and AME in Section~\ref{sec:ame}; we discuss the implications of our results in Section~\ref{sec:discussion} and summarize our principal conclusions in Section~\ref{sec:conclusion}.

\section{Data}
\label{sec:data}
In this section we describe the principal data products employed in this work. Most analysis is done with HEALPix\footnote{\url{http://healpix.sourceforge.net}} maps pixellated with $N_{\rm side} = 128$, corresponding to a pixel size of 27$^\prime$ \citep{Gorski_2005}. A visual summary of the principal datasets employed is provided in Figure~\ref{f:all_maps}.

\subsection{\texorpdfstring{{\rm \hi} Maps}{HI Maps}}

To trace \hi\ in the local ISM, we use $21\rm\,cm$ emission data from HI4PI \citep{hi4pi2016}. HI4PI combines data from the Effelsberg Bonn \hi\ Survey \citep[EBHIS;][]{winkel2010, kerp2011, winkel2016} and the Galactic All Sky Survey at the Parkes radio telescope \citep[GASS;][]{mccluregriffiths2009, kalberla2010, kalberlahaud2015} to produce an all-sky survey of $21\rm\,cm$ emission with $16.^{\prime}2$ angular resolution. We employ the full position-position-velocity (PPV) cubes in our analysis. 

\subsection{PAH Maps}

\subsubsection{\texorpdfstring{$f_{\rm PAH}$}{fPAH}} \label{subsec:fpah_map}
The Wide-field Infrared Survey Explorer (WISE) satellite observed the full sky in four bands \citep{Wright2010}. The W3 band, extending from 7--17\,\micron\ and centered near 12\,\micron, covers the PAH emission features at 7.7, 8.6, 11.3, 12.0, 12.7, 13.55, and 17\,\micron, making it an excellent tracer of Galactic PAH emission. 

In addition to PAH emission, the observed MIR radiation may also arise from starlight and from Zodiacal dust emission. To isolate the PAH component, \citet{Hensley_2016} correlated the diffuse, point-source subtracted W3 emission \citep{Meisner_2014} with the dust radiance $\mathcal{R}$ determined by \citet{Planck_2013_XI} from FIR IRAS and Planck data. Given the relatively high angular resolution of both the W3 and the radiance maps, each $N_{\rm side} = 128$ was subdivided into 120 subpixels each with an independent measurement of the W3 intensity and the dust radiance. A linear fit was performed to determine the slope $f_{\rm PAH}$ in each $N_{\rm side} = 128$ pixel $i$, i.e.,

\begin{equation}
    \left(\nu I_\nu\right)^{12\,\micron}_i = \mathcal{R}_i \times f_{\rm PAH} + C_i
    ~~~,
\end{equation}
where $\left(\nu I_\nu\right)^{12\,\micron}$ is the W3 map after smoothing from the native 15\arcsec\ resolution to 5\arcmin\ resolution and high-pass filtering at 15\arcmin, and $\mathcal{R}$ is the radiance map with a native resolution of 5\arcmin\ high-pass filtered at 15\arcmin. The intercept $C$ absorbs any zero point offsets between the two maps.

With this construction of $f_{\rm PAH}$, the correlation analysis ensures that the PAH component of the W3 emission is isolated, while the normalization by the radiance roughly divides out any dependence on the interstellar radiation field. Thus, $f_{\rm PAH}$ is expected to be proportional to the mass fraction of dust in PAHs. We employ the $N_{\rm side} = 128$ map\footnote{\url{https://doi.org/10.7910/DVN/BJVSNZ}} of $f_{\rm PAH}$ of \citet{Hensley_2016}.

Following \citet{Hensley_2016}, we only employ the $f_{\rm PAH}$ map in pixels where the Pearson correlation coefficient between the W3 intensity and the dust radiance is greater than 0.6. In addition to removing regions where the W3 emission may not be dominated by interstellar dust, this cut also eliminates regions affected by WISE data artifacts, such as the prominent striping from Moon contamination.

\subsubsection{\texorpdfstring{$q_{\rm PAH}$}{qPAH}}

While $f_{\rm PAH}$ has the virtue of being directly determined from the data, it cannot encapsulate the detailed physics of the stochastic heating of dust grains and thus can only serve as a rough proxy for PAH abundance. In contrast, the \citet{Draine+Li_2007} dust model implements detailed calculations of dust temperature distributions in radiation fields of various strengths, enabling a self-consistent calculation of the relative abundances of PAH and other interstellar grains given a dust SED. \citet{Planck_Int_XXIX} fit the \citet{Draine+Li_2007} model to a combination of WISE, IRAS, and Planck data to produce all-sky maps of the fraction of the dust mass in PAHs, $q_{\rm PAH}$.

We employ the $N_{\rm side}=2048$ map\footnote{COM\_CompMap\_Dust-DL07-Parameters\_2048\_R2.00.fits} of $q_{\rm PAH}$ available from the Planck Legacy Archive\footnote{\url{https://pla.esac.esa.int/}}. To downgrade the map to a resolution of $1^\circ$ and $N_{\rm side} = 128$ pixellization, we first use the companion map of the dust mass surface density $\Sigma_d$ to construct a map of the PAH mass surface density $\Sigma_{\rm PAH} = q_{\rm PAH}\Sigma_d$. We then smooth the $\Sigma_d$ and $\Sigma_{\rm PAH}$ maps to 1$^\circ$ resolution and then downgrade each to $N_{\rm side}=128$. Finally, we take the ratio of the smoothed and downgraded $\Sigma_{\rm PAH}$ and $\Sigma_d$ maps to arrive at our final map of $q_{\rm PAH}$.

Comparing \fpah\ and \qpah, we find that the two maps are positively correlated, but have a Spearman rank coefficient of only 0.35 over the area studied in this work. A key limitation of the \qpah\ map is susceptibility to contamination from fluctuations in the cosmic infrared background (CIB), which are difficult to separate from Galactic dust emission. This effect is most pronounced in the high latitude sky where the Galactic emission is the faintest, precisely the region of interest in our study. Whereas \qpah\ is derived from a fit to a dust SED that might include CIB contributions, \fpah\ measures only the correlation between emission in W3 and the dust radiance through a linear fit over a larger sky area. We employ both maps in our analyses to assess the robustness of our conclusions to this uncertainty.

\subsection{AME Maps}

The AME is difficult to isolate from other sources of low-frequency emission, such as free-free, synchrotron, and thermal vibrational dust emission, which are typically of comparable brightness even near the AME peak frequency \citep[e.g.,][]{Planck_Int_XV,Planck_2015_X}. We employ the AME maps\footnote{COM\_CompMap\_AME-commander\_0256\_R2.00.fits} derived from parametric component separation of the microwave sky using the \texttt{Commander} framework \citep{Planck_2015_X}.

The \citet{Planck_2015_X} AME model combines two theoretical spinning dust spectra computed with the SpDust2 software \citep{AliHaimoud_2009,Silsbee_2011} that are then both translated in frequency and scaled in amplitude. The model has three free parameters in each pixel: the amplitude of the first component, the peak frequency of the first component, and the amplitude of the second component. A fourth parameter, the peak frequency of the second component, is fit as a constant value over the full map. We adopt the best fit value of 33.35\,GHz for this latter parameter.

For the remaining parameters, we use the posterior mean values of the \texttt{Commander} fits provided on the Planck Legacy Archive at 1$^\circ$ resolution and pixellated with $N_{\rm side} = 256$. We construct maps of the total AME intensity at 20 and 30\,GHz, which we downgrade to $N_{\rm side} = 128$. We also define an emission-weighted frequency $\langle \nu \rangle_{\rm AME}$ in each pixel as

\begin{equation} \label{eq:nu_ame}
    \langle \nu \rangle_{\rm AME} \equiv \frac{\int \nu I_\nu^{\rm AME}\,d\nu}{\int I_\nu^{\rm AME}\,d\nu}
    ~~~,
\end{equation}
where $I_\nu^{\rm AME}$ is total AME intensity summed over both components in the \texttt{Commander} model. To construct a map of $\langle \nu \rangle_{\rm AME}$ at $N_{\rm side} = 128$, we compute both the numerator and denominator of Equation~(\ref{eq:nu_ame}) at the native $N_{\rm side} = 256$, downgrade each to $N_{\rm side} = 128$, and then take the ratio of the two maps.

\subsection{Dust Emission Map}
Both 12\,\micron\ emission from PAHs and microwave AME are highly correlated with the FIR dust continuum, and so it is convenient to have a representative map of FIR Galactic dust emission. We use the 353\,GHz map\footnote{COM\_CompMap\_Dust-GNILC-F353\_2048\_R2.00.fits} produced using the generalized needlet internal linear combination (GNILC) algorithm to separate Galactic emission from the CIB \citep{Planck_Int_XLVIII}. Following \citet{Planck_2018_XII}, we subtract a CIB monopole of 452\,$\mu$K$_{\rm CMB}$ from the GNILC Stokes I map, then add corrections of 36 and 27\,$\mu$K$_{\rm CMB}$ to account for the Galactic zero level offset based on correlations with \hi\ and the warm ionized medium, respectively. We convert between K$_{\rm CMB}$ and MJy\,sr$^{-1}$ using a conversion factor of 287.5\,MJy\,sr$^{-1}$\,K$_{\rm CMB}^{-1}$ \citep{Planck_2018_III}. Finally, we smooth this map to $1^\circ$ resolution and repixellate to $N_{\rm side} = 128$.

\subsection{Masks}
\begin{figure*}
    \centering
    \includegraphics[width=\textwidth]{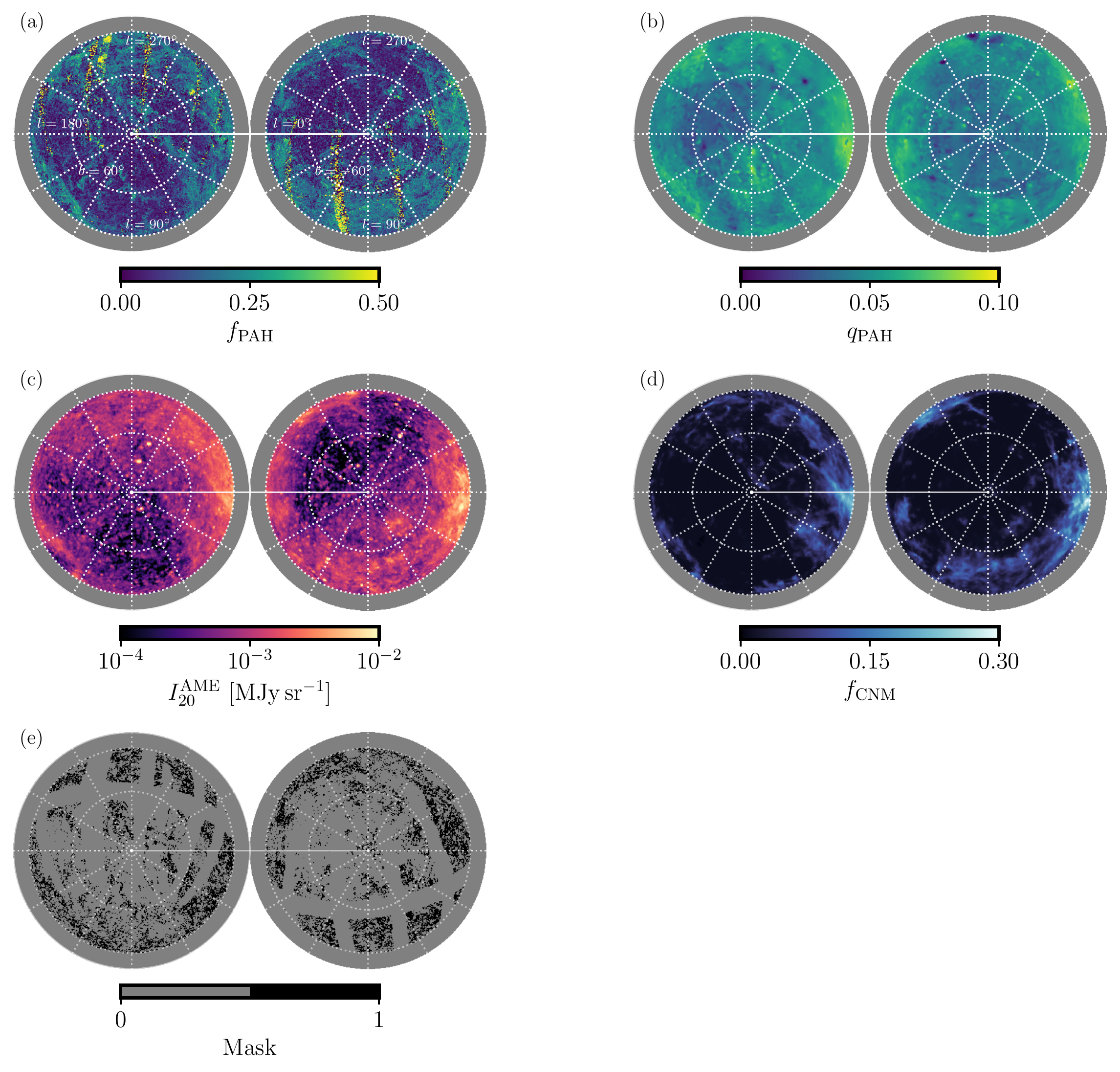}
    \caption{Summary of the key datasets used in this study. Top left (a): $f_{\rm PAH}$ constructed from correlation W3 emission with the FIR dust radiance \citep{Hensley_2016}; top right (b): $q_{\rm PAH}$ as estimated from fits of the \citet{Draine+Li_2007} dust model to WISE, IRAS, and Planck data \citep{Planck_Int_XXIX}; middle left (c): \texttt{Commander} AME map \citep{Planck_2015_X} evaluated at 20\,GHz; middle right (d): $f_{\rm CNM}$ map derived in this work; bottom left (e): mask from \citet{Hensley_2016} used in all analysis in this work (i.e., nonzero pixel values are included in the analysis, zero-valued pixels are excluded).}
    \label{f:all_maps}
\end{figure*}

We employ the mask derived by \citet{Hensley_2016} for all analysis. This mask removes all pixels in which the Pearson correlation coefficient between W3 emission and dust radiance over 120 subpixels is less than 0.6 (see Section~\ref{subsec:fpah_map}). It additionally masks pixels containing bright point sources, artifacts in the WISE map from Moon contamination, all pixels within $5^\circ$ of the ecliptic plane, the pixels in the Galactic plane that constitute the brightest 1\% of the sky, and the 2\% of the sky not observed by IRAS. While the latter cut is unnecessary for our study, the impacts of this cut on our analysis are negligible, so we retain it for consistency with \citet{Hensley_2016}. Finally, as our $f_{\rm CNM}$ map is valid only at Galactic latitudes $|b| > 30^\circ$ (see Section~\ref{sec:fcnm}), we mask this portion of the sky in all analysis. This leaves 9.4\% of the sky unmasked, which we illustrate in Figure~\ref{f:all_maps}.

\section{\texorpdfstring{An $\MakeLowercase{f}_{\rm CNM}$ Map From Full-Sky \hi\ Data}{A Full-sky fCNM Map}}
\label{sec:fcnm}

To trace \fcnm, we apply the CNN model from \citet{murray2020} to HI4PI. This model was trained and tested using augmented, synthetic $21\rm\,cm$ spectra from 3D Galactic ISM simulations \citep{kim2013,kim2014}, and validated using available, sensitive $21\rm\,cm$ absorption observations at high Galactic latitude \citep{murray2020}. From the all-sky HI4PI cubes, we select local Milky Way velocities ($-90<v_{\rm LSR}<90\rm\,km\,s^{-1}$) and resample the data from the native $1.3\rm\,km\,s^{-1}$ per channel velocity spacing to $0.42\rm\,km\,s^{-1}$ per channel velocity spacing to match the inputs to the CNN. Using the public training data and software provided by \citet{murray2020}, following their Section~3.2, we re-train the CNN 25 times and compute \fcnm\ from the input HI4PI cubes at each iteration. In addition to \fcnm\, the CNN outputs a prediction for \rhi, the correction to the \hi\ column density in the optically-thin limit ($N_{\rm \hi}$) due to optical depth. The final maps of \fcnm\ and \rhi\ and their uncertainties are the mean and standard deviation over these trials. We note that this uncertainty is a lower limit, incorporating only the uncertainty in the trained parameter values \citep{murray2020}. Finally, we restrict the resulting all-sky maps to $|b|>30^{\circ}$. 

Where our new HI4PI-based \fcnm\ map overlaps with the map from \citet{murray2020}, we find excellent agreement. After smoothing their GALFA-\hi-based map to $16^{\prime}$ resolution to match HI4PI, we find that the results are positively correlated with Spearman rank correlation coefficient of 0.97. 
 
To match the resolution of the \fpah\ maps, we smooth the \fcnm\ and \rhi\ maps with a Gaussian of FWHM equal to $1^{\circ}$ and resample on a HEALPix grid with $N_{\rm side}=128$. In Figure~\ref{f:all_maps}d we plot the \fcnm\ map. We note that there is an offset in this map of \fcnm$=0.0128$ (and a corresponding offset in the \rhi\ map of \rhi$=1.005$) due to systematic uncertainty in the CNN predictions, and all pixels with \fcnm\ at or below this threshold are considered to be consistent with \fcnm$=0$. 

By inspection of Figure~\ref{f:all_maps}d, the majority of the high-latitude sky features low values of \fcnm. For the unmasked pixels used in this analysis, the median \fcnm$=0.04$. Although the maximum value of \fcnm$=0.3$, only $15\%$ of pixels have \fcnm$>0.1$, and $2\%$ of pixels have \fcnm$>0.2$. This environment is dominated by the WNM.

\section{PAHs and the CNM}
\label{sec:pahs}

\begin{figure*}
\begin{center}
\includegraphics[width=\textwidth]{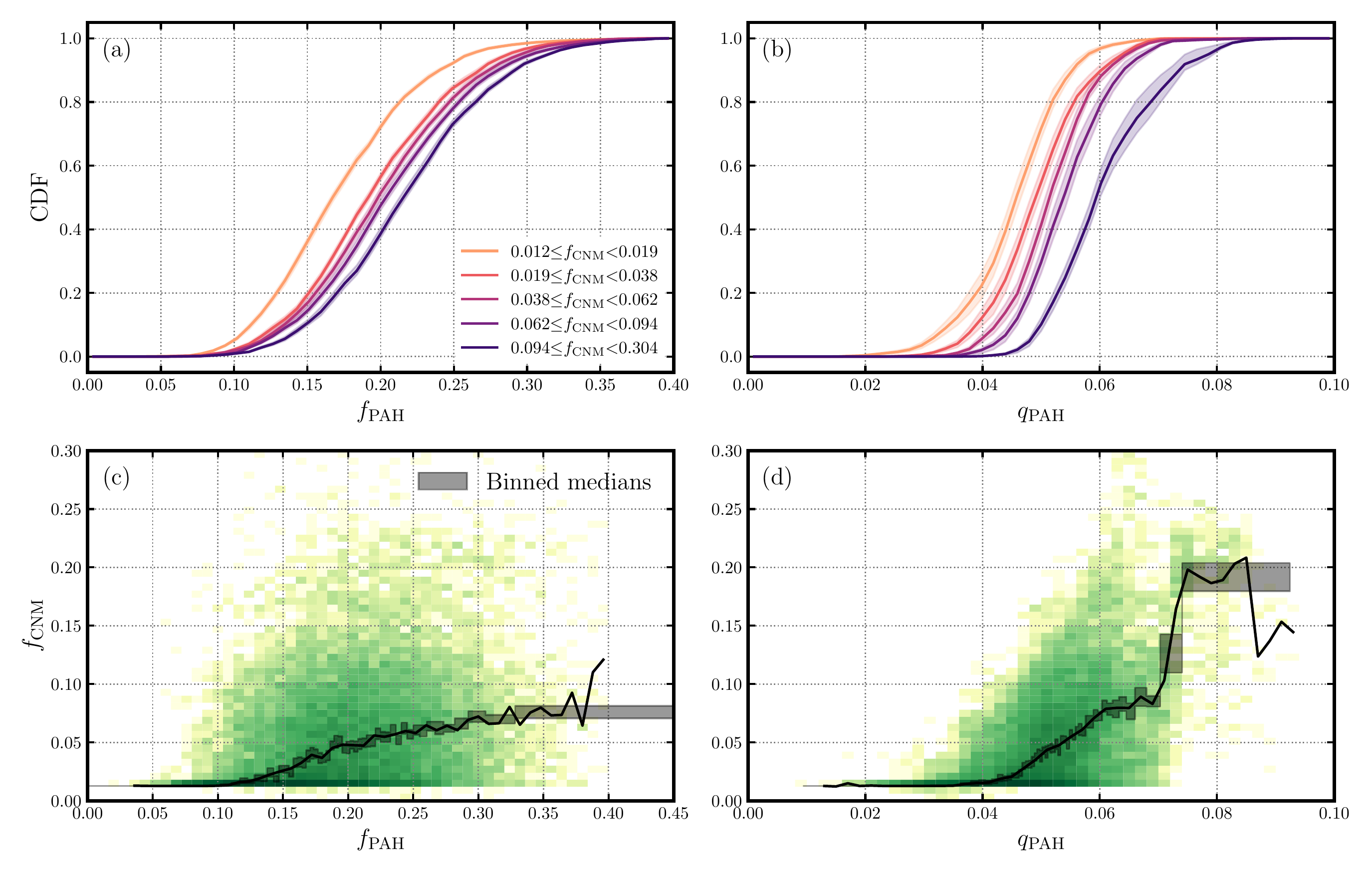}
\caption{Top (panels a, b): Cumulative distribution functions (CDFs) of \fpah\ (a) and \qpah\ (b) in bins of increasing \fcnm. The uncertainty ranges are computed by block-bootstrapping the sky in right ascension (see text) and represent the $1^{\rm st}$ through $99^{\rm th}$ confidence intervals. Bottom: 2D histograms of \fpah\ (c) and \qpah\ (d) vs. \fcnm. In both panels, the median values and uncertainties of \fcnm\ in bins of increasing \fpah\ and \qpah\ computed via block-bootstrapping are shown in gray shading.}
\label{f:fcnm_cdf}
\end{center}
\end{figure*}

To test the relationship between PAHs and the CNM, in Figure~\ref{f:fcnm_cdf} we compare cumulative distribution functions (CDFs) of \fpah\ and \qpah\ in bins of increasing \fcnm. The \fcnm\ limits are selected to produce six equal-size bins for $f_{\rm CNM}>0.0128$ (resulting in $\sim 13600$ pixels per bin). The uncertainties on the CDFs are computed by splitting the sky into 20 equally-spaced bins in right ascension and bootstrapping these bins with replacement over $10^4$ trials. This technique, known as block-bootstrapping \citep{Hall_1985}, accounts for the fact that individual lines of sight are not independent due to the presence of large-angular-scale interstellar structures in the ISM. The shaded uncertainty ranges around each CDF represent the $16\rm th$ through $84\rm th$ percentiles of the block-bootstrapped distributions. 

Figure~\ref{f:fcnm_cdf} demonstrates that \fcnm\ is positively correlated with both \fpah\ and \qpah\ with high significance. This result is insensitive to the choice of block-bootstrapping parameters, or to the bin range selections for \fcnm. In the bottom panels of Figure~\ref{f:fcnm_cdf} we plot 2D histograms of \fpah\ and \qpah\ vs. \fcnm\ to illustrate the parameter distributions in detail, along with running medians. Uncertainties on the running median values are computed by block-bootstrapping within bins of increasing \fpah\ and \qpah, and are overlaid in Figure~\ref{f:fcnm_cdf}c and d. These panels illustrate a qualitatively similar result -- as \fcnm\ increases, so does the PAH fraction as traced by either \fpah\ or \qpah.

We interpret the positive correlation between \fpah\ and \fcnm\ as evidence for a greater PAH abundance per H atom in the CNM relative to the WNM. This may be due to the destruction of PAHs in photoionized gas, which then cools into the WNM, leaving it depleted in PAHs. It is unlikely that the high latitude sightlines probed here have significant reservoirs of dense gas where PAHs could form in situ in the gas phase, but a small amount of high density gas correlated with the CNM could afford some shielding from UV photons and thus reduce destruction rates.

While an increased PAH abundance per H atom in the CNM is the most straightforward explanation of the positive correlation, it is also possible that changes in the PAH charge distribution play a role. Theoretical modeling suggests an increased fraction of ionized PAHs in the WNM relative to the CNM \citep{Weingartner2001}, and the relative strengths of the MIR PAH emission features are sensitive to ionization \citep{Tielens_2008,Maragkoudakis2020,Draine_2021}. However, the W3 passband is broad, and the effects of ionization on individual feature strengths are largely averaged out when integrating over all of the features contributing to the observed W3 emission. We find that the ``low'' PAH ionization model of \citet{Draine_2021} produces only 7\% less emission in the W3 band than the ``high'' ionization model using the fiducial grain size distribution and radiation field parameters from that study. Thus, it is unlikely that the effects of grain charge alone can explain the observed trend of increasing \fpah\ with increasing \fcnm.

\section{AME and the CNM}
\label{sec:ame}

\begin{figure*}
\begin{center}
    \includegraphics[width=\textwidth]{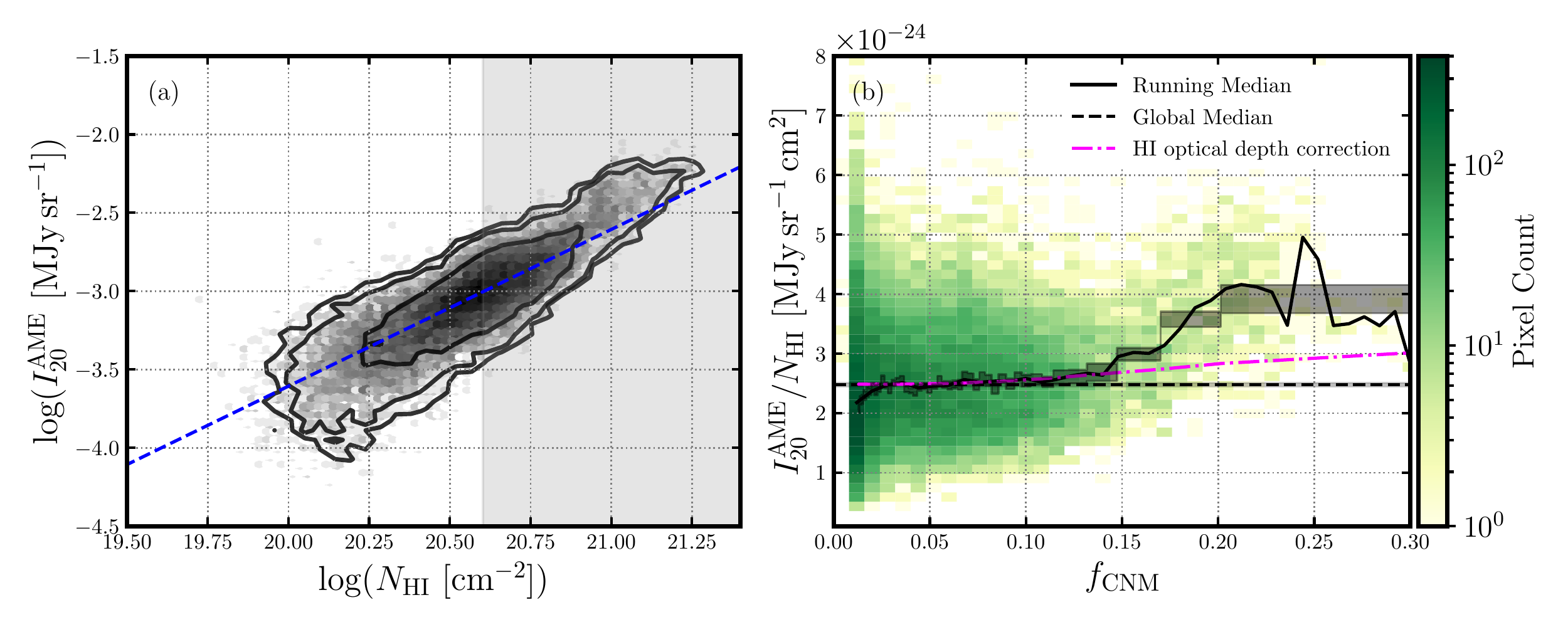}
    \caption{(a): 2D histogram of the 20\,GHz AME versus the \hi column density, displaying a strong, approximately linear correlation between the two quantities. The median value of $I_{20}^{\rm AME}/N_{\rm \hi}$ for all pixels with $N_{\rm \hi}<4\times10^{20}\rm\,cm^{-2}$ is indicated in the blue dashed line. (b): 2D histogram of 20\,GHz AME per $N_{\rm \hi}$ versus $f_{\rm CNM}$. To compare with the global median (blue dashed), we plot the median $I_{20}^{\rm AME}/N_{\rm \hi}$ in bins of increasing \fcnm\ with uncertainties computed via block-bootstrapping (black lines, gray shading) and find a significant positive correlation, exceeding the correction to the global median due to \hi\ optical depth (magenta dot-dashed line).}
    \label{f:hi_ame}
\end{center}
\end{figure*}

\begin{figure*}
\begin{center}
    \includegraphics[width=\textwidth]{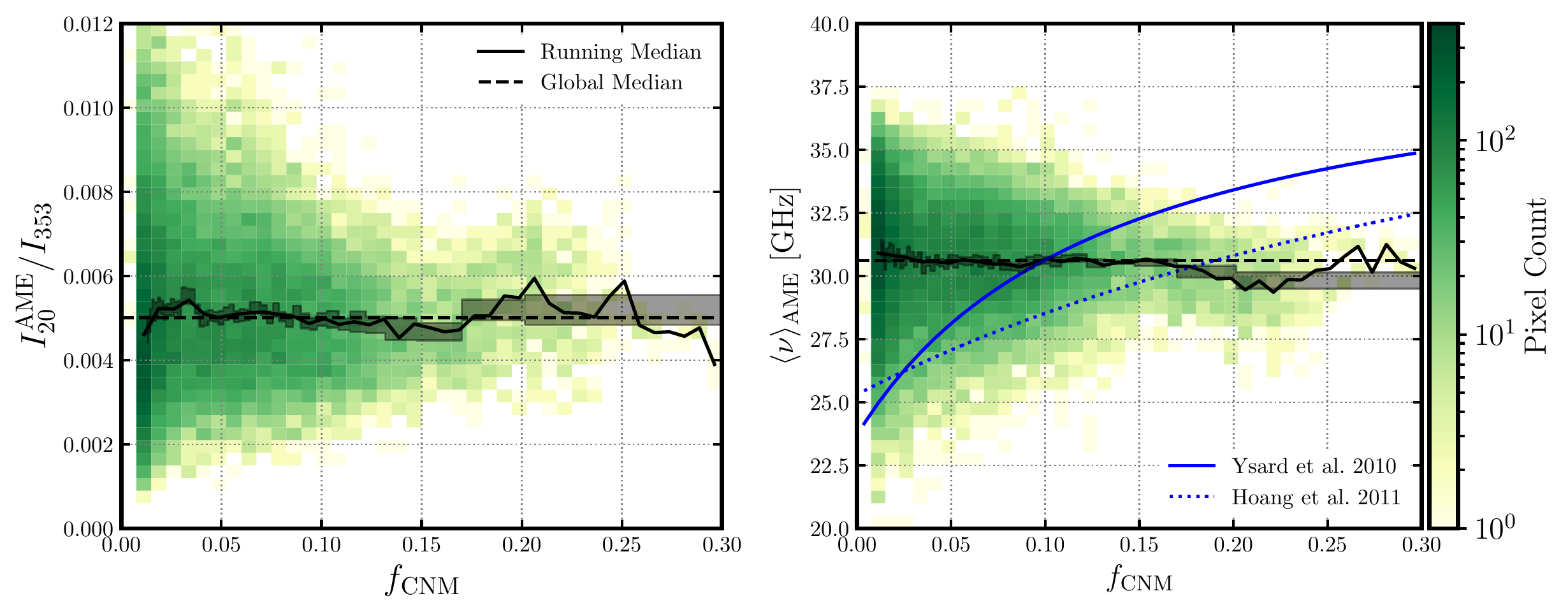}
    \caption{2D histograms of the 20\,GHz AME per 353\,GHz dust emission ($I_{20}^{\rm AME}/I_{353}$, left) and the AME-weighted frequency ($\langle \nu \rangle_{\rm AME}$, right) versus the CNM fraction $f_{\rm CNM}$. Uncertainties in the running medians (black lines) are computed via block-bootstrapping (gray shading). No significant correlations are found. In the right panel, we compare two predictions based on the models of \citet{Ysard_2010b} and \citet{Hoang_2011}, finding that both models predict stronger dependence on \fcnm\ than is observed.}
    \label{f:fcnm_ame}
\end{center}
\end{figure*}

It has been shown that, even in the low column density regime we focus on in the present study, the FIR emission from interstellar dust per H atom increases monotonically with $f_{\rm CNM}$ \citep{Clark_2019,murray2020}. It is unclear whether this indicates a higher dust-to-gas ratio in the CNM relative to the WNM, a larger emissivity of grains in the CNM, or the presence of molecular gas correlated with the CNM. In the previous section, we demonstrated that PAH emission has an even stronger positive correlation with $f_{\rm CNM}$ than the FIR continuum. We thus first investigate whether this is true for AME as well.

Figure~\ref{f:hi_ame}a presents the correlation of the 20\,GHz AME with $N_{\rm \hi}$, demonstrating strong positive correlation. This is not unexpected, as AME is known to correlate well with FIR dust emission \citep{Kogut_1996,Finkbeiner_2004,Hensley_2016,Dickinson_2018}, which is itself strongly correlated with $N_{\rm \hi}$ over this range of column densities \citep[$N_{\rm \hi} < 4\times10^{20}$\,cm$^{-2}$][]{Lenz_2017}. In Figure~\ref{f:hi_ame}b, we plot the 20\,GHz AME per $N_\hi$ as a function of $f_{\rm CNM}$. We compare the global median $I_{20}^{\rm AME}/N_{\rm \hi}$ with median values in bins of increasing \fcnm, finding a clear positive correlation. The positive trend exceeds the prediction due to the effects of \hi\ optical depth, computed by correcting the $I_{20}^{\rm AME}/N_{\rm \hi}$ ratio by \rhi. Thus, like the FIR continuum and like the 12\,\micron\ PAH emission, there is more AME per H atom when $f_{\rm CNM}$ is larger.

If PAHs are the carriers of the AME, then, all else equal, an enhanced abundance of PAHs implies more AME. However, \citet{Hensley_2016} found no evidence for a correlation between the AME per unit FIR dust emission and emission in the WISE 12\,\micron\ emission per unit FIR dust emission. To investigate correlations between AME and the CNM, we plot the 20\,GHz AME intensity ($I_{20}^{\rm AME}$) per 353\,GHz dust intensity ($I_{353}$) as a function of \fcnm. We find no compelling evidence for a correlation. At the highest $f_{\rm CNM}$ values in the map ($f_{\rm CNM} \simeq 0.2$), there is a slight tendency for more AME per $I_{353}$. However, this fluctuation is not significant relative to the sample variance as estimated by block bootstrapping, as the uncertainty bands in Figure~\ref{f:fcnm_ame} indicate. Thus, unlike for PAHs, we find no evidence for enhanced AME relative to FIR continuum emission as $f_{\rm CNM}$ increases. Qualitatively similar results are obtained using the map of AME at 30\,GHz.

A prediction of several spinning dust models in the literature \citep[e.g.,][]{Draine_1998,Ysard_2010a,Hoang_2011} is that AME associated with the WNM peaks at lower frequencies than does AME from the CNM. To investigate whether such a correlation is present in our data, we correlate the emission-weighted frequency $\langle \nu \rangle_{\rm AME}$ (Equation~(\ref{eq:nu_ame})) with $f_{\rm CNM}$. As illustrated in the right panel of Figure~\ref{f:fcnm_ame}, there is no compelling evidence for a correlation. As with $I_{20}^{\rm AME}/I_{353}$, there is a suggestion of qualitatively different behavior around $f_{\rm PAH} \simeq 0.2$ where $\langle \nu \rangle_{\rm AME}$ appears to be slightly lower than average, but we again interpret this with caution given the large sample variance and the limited sky area over which pixels with such high $f_{\rm PAH}$ values are found.

To compare this null result to theoretical predictions, we estimate $\langle \nu \rangle_{\rm AME}$ for two models. \citet{Ysard_2010b} present spectra of the AME in the CNM and WNM which fit radio observations when $f_{\rm CNM} = 0.1$. We compute the combined spectrum at all values of $f_{\rm PAH}$ by rescaling the WNM and CNM components of their model appropriately and then compute $\langle \nu \rangle_{\rm AME}$. We repeat this procedure for the model of \citet{Hoang_2011}, who employ $f_{\rm CNM} = 0.2$ to fit a mean AME spectrum. As expected, both models agree well with the global median $\langle \nu \rangle_{\rm AME}$ when $f_{\rm CNM}$ is equal to the value adopted in each study. However, Figure~\ref{f:fcnm_ame} demonstrates that neither of these models are a good representation of the data, predicting far more dependence on \fcnm\ than is observed.

\section{Discussion}
\label{sec:discussion}

Both PAH emission in the W3 band and the AME are highly correlated with the FIR continuum. We have demonstrated, however, that these correlations depend differently on $f_{\rm CNM}$. We find that the PAH emission per FIR continuum emission, as quantified by $f_{\rm PAH}$ and $q_{\rm PAH}$, increases with increasing $f_{\rm CNM}$. We posit that grain destruction mechanisms, particularly photodestruction by UV radiation, lead to a reduction in the number of PAHs per H atom in the WNM relative to the CNM.

On the other hand, we find that the AME per FIR continuum emission has little correlation with $f_{\rm CNM}$. Models of the observed AME SED in the diffuse ISM have generally invoked both a CNM and WNM component to explain its observed width, with PAHs in the WNM dominating at low frequencies and PAHs in the CNM dominating at high frequencies \citep{Ysard_2010b,Hoang_2011}. Parametric fits using the Bayesian \texttt{Commander} framework have likewise required two distinct components, though they are not explicitly identified with the CNM and WNM \citep{Planck_2015_X}. In all cases, the low frequency component dominates the emission by a factor of a few. We find very little dependence of the AME spectrum on $f_{\rm CNM}$ (Figure~\ref{f:fcnm_ame}) and that such models strongly overpredict changes in the AME SED with changes in $f_{\rm CNM}$.

If PAHs are the AME carriers and if the PAH abundance per H atom is reduced in the WNM relative to the CNM, then it must be the case that PAHs in the WNM are better able to produce AME than PAHs in the CNM. Given the lack of evidence for different AME SEDs in the WNM versus CNM (Figure~\ref{f:fcnm_ame}), the emitting grains must have similar rotational frequencies. AME is typically dominated by the smallest grains able to survive in the interstellar radiation field \citep[e.g.,][]{Hensley_2017}, and so one possibility is that the WNM has a greater abundance of ultrasmall grains able to spin at 20--30\,GHz frequencies while the size distribution of PAHs in the CNM has relatively more grains at slightly larger sizes that cannot emit at AME frequencies but still produce thermal emission at 12\,\micron. Another possibility is systematic differences in the electric dipole moments of PAHs in the WNM vs CNM, which can alter both the total emission per grain as well as the grain rotational velocity distribution. If PAHs in the WNM have a comparable rotational velocity distribution but a larger electric dipole moment, then the AME would be enhanced relative to the CNM. A third possibility is differences in the grain charge distribution. Particularly in the CNM, Coulomb focusing of gas cations results in enhanced rotational excitation of negatively charged grains \citep{Hensley_2017}. If theoretical calculations have overpredicted the abundance of negatively charged PAHs in the CNM and thus the importance of this effect, the predicted AME spectra of the WNM and CNM would be much more similar.

Ultimately, the results presented here suggest that the lack of correlation between the fraction of dust in PAHs and the AME per FIR emission may not be evidence that PAHs are not the AME carriers. Rather, it may simply be that PAH emission per unit FIR continuum is somewhat larger in the CNM than in the WNM whereas the AME per unit FIR continuum is somewhat less due to the excitation and destruction mechanisms that differ between these environments.

A potential test of this hypothesis is use of the 3.3\,\micron\ PAH feature as a proxy for the abundance of the smallest grains. If the WNM has relatively more of the smallest PAHs than the CNM, then the 3.3\,\micron\ emission should originate preferentially from the WNM. All sky maps of DIRBE 3.5\,\micron\ emission include this feature and could be used to test this prediction.

We emphasize that this study has focused on PAH emission and AME from the diffuse ISM with $N_{\rm \hi} < 4\times10^{20}$\,cm$^{-2}$. Correlations in dense cloud environments are expected to be qualitatively different than those found here, on account of, e.g., PAHs sticking onto dust grains at high densities, as well as the conversion of PAHs to other molecular species such as fullerenes. Differences in the PAH size distribution and excitation mechanisms are likely even more important to consider in molecular clouds than in diffuse gas, and the qualitative differences between correlations in clouds versus the diffuse ISM that have been observed \citep[e.g.,][]{ArceTord_2020} are perhaps not unexpected.

The fidelity of our study is likely most limited by the reliability of the AME maps, which result from an extensive and challenging component separation analysis \citep{Planck_2015_X}. While additional multifrequency data have in general validated the \citet{Planck_2015_X} AME maps, systematic discrepancies have been noted at the $\sim$10--20\% level \citep{Dickinson_2019,Poidevin_2019,CepedaArroita_2021}. In particular, the AME is often underestimated and free-free emission overestimated. While we have performed various robustness tests of our results, improving the AME component via incorporation of additional low frequency data sets such as QUIJOTE \citep{RubinoMartin_2012}, C-BASS \citep{Jones_2018}, and S-PASS \citep{Carretti_2019} is critical for further detailed study of the AME spectrum and its dependence on interstellar environment.

In addition, we are limited by the CNN to probe only diffuse \hi\ gas. Extending this study to lower Galactic latitudes will require a new \fcnm\ model, trained with synthetic observations of next-generation numerical simulations with sufficient resolution to resolve the CNM in dense environments. Large-area surveys of $21\rm\,cm$ absorption planned for the Square Kilometer Array and its pathfinders \citep[e.g.,][]{dickey2013, mcclure-griffith2015} will significantly expand the available data needed to validate new CNNs to properly account for the effects of self-absorption and velocity blending of \hi\ lines. With these in hand, we look forward to analyzing the relationships between \fcnm, PAHs, and the AME in denser, dustier environments. 

\section{Conclusions}
\label{sec:conclusion}

We have presented a new map of the fraction of Galactic \hi\ in the CNM, $f_{\rm CNM}$, over the full high Galactic latitude sky. Using this $f_{\rm CNM}$ map, we have investigated the relationships among PAH emission, the AME, and the CNM on diffuse lines of sight. The principal conclusions of this work are as follows:

\begin{enumerate}
    \item We establish that the fraction of dust in PAHs increases with increasing $f_{\rm CNM}$. We interpret this as evidence of PAH destruction in warm gas.
    \item AME per $N_{\rm \hi}$ increases with increasing $f_{\rm CNM}$, just as FIR dust emission does. On the other hand, we find little correlation between AME per unit dust emission and $f_{\rm CNM}$. This implies that AME and 12\,\micron\ PAH emission do not have the same emissivity per H atom in different interstellar environments.
    \item We find no evidence for changes in the AME SED, in particular its peak frequency, with $f_{\rm CNM}$. This is contrary to expectations from models which interpret the observed AME spectrum as a combination of CNM and WNM components peaking at higher and lower frequencies, respectively.
    \item We suggest that the weaker than expected correlation between 12\,\micron\ PAH emission and the AME might be explained by the different grain properties (e.g., size, charge, electric dipole moment) and excitation mechanisms of PAHs in the WNM versus CNM.
\end{enumerate}

Detailed modeling is required to assess whether the changes in 12\,\micron\ and 20\,GHz emissivities between the WNM and CNM due to, e.g., changes in the grain size distribution are sufficient to explain the observed trends. Nevertheless, this study makes the case on empirical grounds that such changes may underlie the weaker than expected correlation between the PAH abundance and the AME and thus that PAHs may indeed be the AME carriers after all.

\begin{acknowledgments}
We thank Bruce Draine, Joshua E.\,G. Peek, and Marc-Antoine Miville-Desch{\^e}nes for useful discussions and helpful feedback. BSH acknowledges support from the NASA TCAN grant No. NNH17ZDA001N-TCAN. CEM acknowledges support from the NSF Astronomy and Astrophysics Postdoctoral Fellowship under award AST-1801471. MD acknowledges support from the Princeton Undergraduate Summer Research Program. The authors acknowledge the program ``The Grand Cascade'' hosted by Institut Pascal at the Universit\'{e} Paris-Saclay and the Interstellar Institute for fostering discussions which greatly improved this work. Some of the results in this paper have been derived using the healpy and HEALPix packages.
\end{acknowledgments}

\software{healpy \citep{Gorski_2005,Zonca_2019}, Matplotlib \citep{Matplotlib}, NumPy \citep{NumPy}}

\bibliography{refs}
\end{document}